\documentclass[a4paper, notoc]{JHEP3}
\usepackage[dvips]{graphicx}
\usepackage{amsfonts}
\usepackage{amssymb}
\usepackage{epsfig}
\usepackage{cite}

\newcommand{\be}{\begin{equation}}
\newcommand{\ee}{\end{equation}}
\newcommand{\bea}{\begin{eqnarray}}
\newcommand{\eea}{\end{eqnarray}}
\newcommand{\mbb}{\mathbb}
\newcommand{\ti}{\times}
\newcommand{\half}{\frac{1}{2}}
\newcommand{\mc}{\mathcal}

\newcommand{\beqa}{\begin{eqnarray}}
\newcommand{\eeqa}{\end{eqnarray}}

 \newlength{\wth}
 \newcommand{\twographs}[2]{%
 \unitlength=1.1in
 \begin{picture}(5.8,2)
 \put(0,0){\epsfig{file=#1.eps, width=0.7\wth}}
 \put(2.5,0){\epsfig{file=#2.eps, width=0.7\wth}}
 \put(0,1.7){(a)}
 \put(2.5,1.7){(b)}
 \end{picture}
}

 \title{Volume  Modulus Inflation and the Gravitino Mass Problem}
\author{J.P. Conlon$^{1,2}$, R. Kallosh$^{3,4}$, A. Linde$^{3,4}$ and F. Quevedo$^{2}$
 \\$^{1}$Cavendish Laboratory, J J Thomson Avenue,
 Cambridge CB3 0HE, UK \\
 \\$^{2}$DAMTP,
  Centre for Mathematical
Sciences,\\
  Wilberforce Road, Cambridge, CB3 0WA, United Kingdom\\
\\
$^{3}$Department of Physics, Stanford University,
Stanford, CA 94305, USA\\
\\
${}^4$ Yukawa Institute for Theoretical Physics, Kyoto, Japan \\
}

\abstract{The Hubble constant during the last stages of inflation in a broad class of models based on the KKLT mechanism should be smaller than the gravitino mass,  $H \lesssim m_{3/2}$. We point out that in the models with large volume of compactification the corresponding constraint typically is even stronger, $H \lesssim m_{3/2}^{3/2}$, in Planck units. In order to address this problem, we propose  a class of models with large volume of compactification where inflation may occur exponentially far away from the present vacuum state. In these models, the Hubble constant during inflation can be many orders of magnitude greater than the gravitino mass. We introduce a toy model describing this scenario, and discuss its strengths and weaknesses.
}

\preprint{DAMTP-2007-36 \\ CAV-HEP-08/07 \\ YITP-08-42}

\begin{document}

\tableofcontents

\section{Introduction: Inflation and the Gravitino Mass}

A realistic model of the universe should be consistent with
all constraints, not just from high energy physics experiments but also from
cosmological observations. Inflation is the leading scenario for the early universe
and low-energy supersymmetry is the leading proposal for new physics subject to
experimental test at the LHC. It is therefore important to develop models
capable of including both these desirable features of physics beyond the standard model.

Many  recent attempts to implement inflation in the context of string theory
are based on the KKLT mechanism of vacuum stabilization \cite{kklt}
and its  generalizations \cite{hepth0502058}. Until very recently, the standard assumption
of many of these inflationary models was that inflation is a very high
energy scale phenomenon, and therefore one can construct inflationary models quite
independently of the requirements of the low-energy SUSY phenomenology.

However,  recent studies of inflationary models in string theory revealed a rather unexpected fact:
in the simplest models based on the KKLT mechanism the Hubble constant $H$  should be  smaller than the present value of the gravitino mass  \cite{kl},
\begin{equation}\label{constr}
 H \lesssim m_{{3/2}} \ .
\end{equation}
The reason for this bound is that the gravitino mass
at the supersymmetric KKLT minimum, with $DW=0$ before the uplifting, is given by
$
3m_{{3/2}}^2= |V_{AdS}|
$.
Uplifting of the AdS minimum to the present nearly Minkowski vacuum
occurs by adding to the potential a term of the form $C/\sigma^{n}$, where $\sigma$ is the volume modulus
and $n=3$ for a generic compactification and $n=2$ for the highly warped throat geometry.
Since the uplifting is less significant at large $\sigma$, the
energy barrier to decompactification created by the uplifting
is generically slightly smaller than $|V_{AdS}|$: \, $V_{{\rm barrier}} \lesssim |V_{AdS}| \sim 3m_{{3/2}}^2$.
However, the volume modulus couples to all sources of energy
due to the Weyl rescaling always present when deriving the 4-dimensional action.
In particular,
the energy of the inflaton field $\phi$ will give an additional uplifting of a similar
type: $\Delta V(\phi,\sigma) \sim V(\phi)/\sigma^{n}$.
As this is also proportional to an inverse power of the volume modulus, it is
larger at the minimum of the KKLT potential than at the top of the barrier.
Therefore adding a large vacuum energy density to the KKLT potential,
as required for inflation, may destabilize the minimum by uplifting it
to a height greater than
the height of the barrier, see Fig. \ref{bound}.
In typical KKLT-type models this leads to vacuum destabilization if the added
energy density $V(\phi)/\sigma^{n}$, which is responsible for inflation, is much greater
than the height of the barrier $V_{{\rm barrier}} \lesssim 3m_{{3/2}}^2 M_P^2$.
Since $H^{2} \sim  \Delta V(\phi,\sigma)/3$, this leads to the bound (\ref{constr})
(see \cite{kl} for a more detailed discussion of this issue, while a similar problem in a
slightly different context was also found in \cite{Buchmuller:2004tz}).

\begin{figure}[h!]
\centering\leavevmode\epsfysize=5.5cm \epsfbox{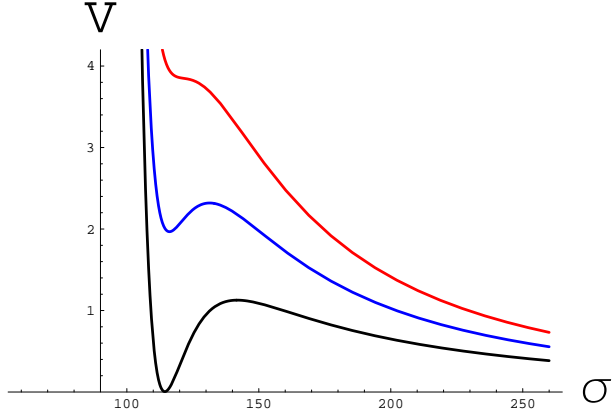} \caption[fig2]
{The lowest curve with dS minimum is the potential of the KKLT model. The second one shows what happens to the volume modulus potential when the inflaton potential $V_{\rm infl}={V(\phi)\over \sigma^3}$ added to the KKLT potential. The top curve shows that when the  inflaton potential becomes too large, the barrier disappears, and the internal space decompactifies. This explains the constraint $H\lesssim   m_{3/2}$.  } \label{2}
\end{figure}\label{bound}

In KKLT-based models, it therefore seems that for a gravitino
mass $m_{3/2} \sim 1 \hbox{TeV}$ the Hubble constant
during the last stages of a string theory inflation model should be quite
low, $H \lesssim 1$ TeV, which is ten orders of magnitude below the often discussed GUT inflation scale.
Therefore if one believes in standard SUSY phenomenology with $m_{{3/2}} \lesssim O(1)$ TeV, one should find a realistic particle physics model where  the nonperturbative string theory dynamics occurs at the LHC scale or even lower (the mass of the volume modulus in the KKLT scenario typically is not much greater than the gravitino mass), and inflation occurs  at a density at least 30 orders of magnitude below the Planck energy density  \cite{kl}.
For a recent analysis of this issue see e.g. \cite{Covi:2008cn} and
for a discussion in the context of the heterotic string see \cite{Lalak:2005hr}.

This problem is quite generic. For example, recently a new interesting mechanism of moduli stabilization was proposed, which is based on the models with compacification on Nil manifolds with negative curvature \cite{Silverstein:2007ac}. This mechanism presents
 a  significant modification of the compactifications on flat Calabi-Yau spaces, as suggested by the assumption of the low scale supersymmetry. And yet, the same constraint $H\lesssim   m_{3/2}$ remains valid for the inflationary models in this scenario \cite{Silverstein:2008sg}.

The situation becomes even trickier in the large volume models of vacuum stabilization  \cite{hepth0502058}.
In such models the height of the barrier is much smaller, $V_{{\rm barrier}}  \sim m_{{3/2}}^3 M_P$.
In this case, the constraint that the inflaton potential should not be much greater than the height of the barrier
leads to the bound  (in  units $M_{p}= 1$)
\be
H \lesssim m_{3/2}^{{3/2}} \ .
\ee
For $m_{3/2} \sim 1$ TeV this inequality implies that the Hubble constant
during inflation in this class of models \cite{hepth0509012,roulette} cannot exceed $O(1)$ KeV,
which is an extremely strong constraint.

There do exist proposals of low-scale inflationary models, for example the so-called MSSM inflation,
which may occur for $H \sim 10$ GeV or even for $H \sim 10$ MeV \cite{MSSM}.
Ref. \cite{rosssarkar} also contains a discussion of models where inflation may
occur at extremely low scales, with an example of a model for which $H \sim 10^{-7}$ eV.
 In particular, if the inflaton potential energy at $H \sim 1$ KeV could instantly transfer to
 thermal energy, the corresponding temperature would be about $10^{6}$ GeV, which is much greater than the critical temperature of the phase transition in the standard model.
If this instantaneous transition is achievable, the
temperatures would then be sufficiently high for the subsequent generation of a baryon asymmetry.

One can find models with a very low-scale inflation in the
context of the KKLT or large volume scenarios, since the energy scale
is exponentially sensitive to the
parameter $a$ of the nonperturbative superpotential $W = W_{0}+ Ae^{-aT}$ \cite{kklt}.
%{\bf Which models does this refer to - is this the racetrack model with the par%ameters rescaled?}
 However, models of this type are very non-traditional, and their parameters
are substantially different from the parameters of all current existing models of string theory inflation.
Furthermore, as the required value of the slow-roll epsilon
parameter is given by $\epsilon \sim (E_{\rm inf} /6 \ti 10^{16} \hbox{GeV})^4$, low-scale inflation
substantially increases the amount of fine-tuning required in the inflaton potential.
It is important to know whether this tension between high-scale inflation and
TeV supersymmetry is unavoidable or whether it is simply a consequence of the assumptions used so far in inflationary
model-building.

This is not the first time that string theory and supergravity have encountered
cosmological problems associated
with the small value of the gravitino mass and of the moduli fields.
The famous gravitino problem and the cosmological moduli problem are haunting us for more than
two decades  \cite{gravitinos,Coughlan:1983ci, cmp}. Now we see that the smallness of the gravitino mass leads to an
additional problem in the context of string cosmology \cite{kl,Buchmuller:2004tz}. This problem would disappear if
one would consider supersymmetric models with large gravitino mass, for example \cite{DeWolfe:2002nn,Arkani-Hamed:2004fb},
or used a solution to the hierarchy problem different to that of TeV
supersymmetry.\footnote{For other problems with high values of the Hubble constant in string inflation
  see \cite{myers}.}

There exist ways to address this problem without increasing the value of the gravitino mass.
For example, one may consider KKLT models with the
racetrack superpotential containing at least two exponents and find
parameters such that the supersymmetric minimum of the potential, even prior to uplifting,
occurs at zero energy density \cite{kl}, which would mean $m_{3/2} = 0$. By a slight change of parameters in this class of models, which are sometimes called KL models,
one can get a gravitino mass that is nonzero but still much smaller than the height of the barrier, removing
the constraint $H \lesssim m_{{3/2}}$. In particular, one can use the KKLMMT brane inflation model \cite{Kachru:2003sx,bcsq,Baumann:2007np,delicate,Krause:2007jk,Panda:2007ie,Itzhaki:2007nk} and implement it in the context of the KL scenario with $H \gg m_{3/2}$.

The difficulty with this solution is that if we want to have $H$ many orders of magnitude greater
than $m_{{3/2}} \sim 1$ TeV, we need to fine-tune the parameters of the model to a corresponding accuracy.
The origin of the electroweak scale is then a kind of accident,
reducing the attractiveness of supersymmetric solutions to the hierarchy problem.
However, this class of models has certain advantages from the point of view of vacuum stabilization \cite{kl}, so
it might happen that the required fine-tuning is not unreasonable.

Another possible solution of this problem was recently proposed in \cite{Badziak:2008yg} in the context of the volume modulus inflation
in the KKLT scenario. This model extended the KL model to involve triple gaugino condensation in the superpotential and also
required a modification of the  K\"ahler potential. It introduces 6 new parameters (two real parameters and two complex ones), and
fine-tuned three of these parameters with accuracy ranging from $10^{-4}$ to $10^{{-7}}$.
This clearly demonstrates that it is quite difficult to avoid the constraint  $H< m_{3/2}$ in such models, but
nevertheless it is encouraging that it is possible to do so.

In this paper we will concentrate on the models with large volume of compactification \cite{hepth0502058} and propose another possible resolution of the gravitino mass problem, aiming at making a transition of scales from $E_{inf}$ to $E_{susy}$ natural.
We shall discuss our idea and certain issues associated with it  in the next sections.

\section{Disentangling $H$ and $m_{{3/2}}$: the basic idea}\label{1fm}

The idea we revisit in this paper is the assumption
used so far in building supergravity models of inflation that the
true minimum of the scalar potential is relatively close in field space to the locus at which
inflation ends.
We instead propose that inflation
should end with a runaway in field space, with the true minimum lying a very long distance in field
space from the location where inflation occurs (with the model discussed below, the distance in canonically normalized
field space will correspond to approximately twenty Planckian distances). The problem we are trying to address
is that characteristic inflationary energy scales
are much larger than those appropriate for supersymmetry breaking.
The advantage of a runaway epoch is that
evolution along runaway potentials (e.g. of the form $V(\phi) \sim V_0 e^{-\lambda \phi}$)
is one of the few efficient ways of naturally dissipating large quantities of energy and reducing the scale of the potential
by many orders of magnitude.
If the true minimum of the scalar potential
lies a long way along the runaway direction, it naturally has much lower characteristic energy scales
than apply during inflation.

\begin{figure}[h!]
\centering{\includegraphics[height=9cm]{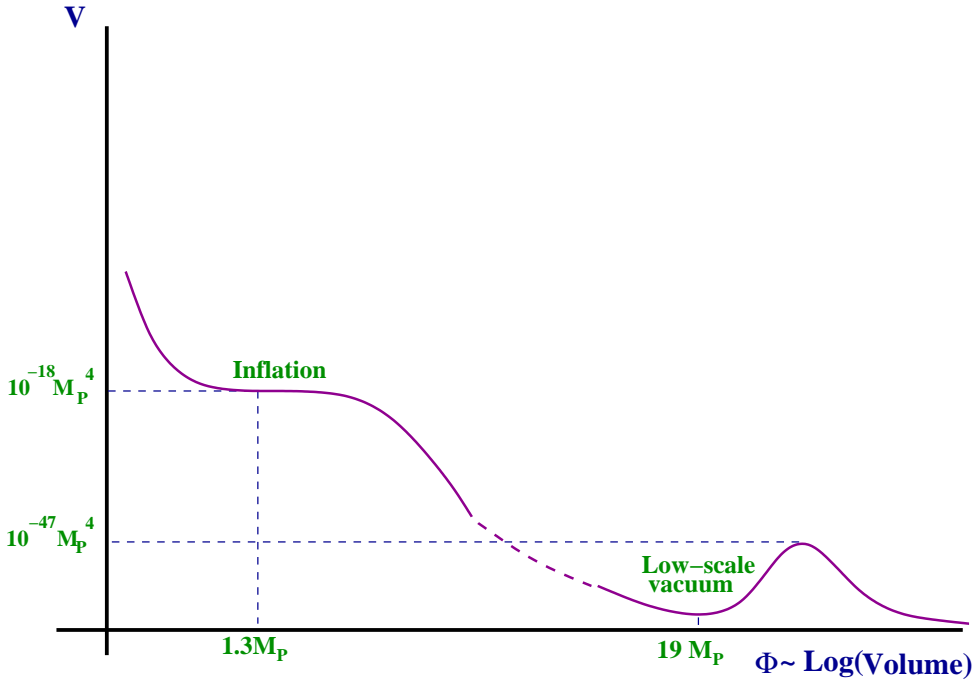}}
 \caption{An illustration of the scenario
        put forward in this article. At relatively small volume,
high-scale inflation occurs due to fine-tuned quantum corrections.
 After inflation the volume modulus evolves over a long range of many Planck scales,
 eventually settling in the large volume minimum with TeV gravitino mass. Although
 the barrier protecting from decompactification is very
 small compared to the initial energies, an attractor solution guides the fields to the minimum
 and prevents overshooting.} \label{illustration}
\end{figure}

In this case supergravity models of high-scale inflation consistent
with low-scale supersymmetry breaking, $m_{3/2} \lesssim 1\ \hbox{TeV}$
\emph{in vacuo}, should have three stages. In the first, inflation occurs at high energy scales with $m_{3/2} \gg 1\  \hbox{TeV}$
during inflation. In the second, inflation ends with the fields fast-rolling
towards a runaway direction. For example, in the model below
this is due to inflation occurring near an inflection point in the volume direction.
As inflation ends it is necessary that trace quantities of radiation be generated to act as
a seed for an attractor solution. In the third stage,
the presence of small initial quantities of radiation drives the fields to an attractor solution.
The attractor solution applies during the runaway epoch and dissipates energy.
The scaling nature of the attractor solution avoids overshooting
and guides the fields
into the global minimum of the potential in which
$m_{3/2} \sim 1\ \hbox{TeV}$.\footnote{For a recent discussion 
in the context of M-theory compactifications of
the overshooting problem and how to avoid it, see \cite{08040863}.} 
This scenario is illustrated in Fig. \ref{illustration}.

The justification for the existence of a minimum at very large values of the volume,
far along the runaway direction, is the large volume scenario
\cite{hepth0502058}, where the inclusion of $\alpha'$ corrections into the KKLT framework generates a new minimum of the
scalar potential at exponentially large values of the volume, with hierarchically small values of $m_{3/2}$.

To illustrate this idea, we start by studying moduli evolution in the following toy model describing a field $\Phi$ with a potential
\begin{equation}
V\ =\ V_0 \left((1-\epsilon \, \Phi^{3/2})\, e^{-\sqrt{\frac{27}{2}}\Phi}\,
+C\, e^{-10\Phi/\sqrt{6}}+ D\, e^{-11\Phi/\sqrt{6}}\, +\, \delta\,
e^{-\sqrt{6}\Phi}\right).
\label{1modpot}
\end{equation}
The particular form of this potential is motivated by that
arising as the effective potential for the volume modulus in the large volume models.
The connection to the large volume models and the supergravity origin of the above potential
will be discussed in the next section. The 1-modulus potential (\ref{1modpot}) will not represent a
complete model but will allow us to capture several key features of our proposal.

In (\ref{1modpot}) $\Phi$ is the canonically normalized volume modulus, $\Phi=\sqrt{3/2}\, \log \tau_b$
with $\mc{V} = \tau_b^{3/2}$, so $\tau_b$ is the volume of a 4-cycle.
The first two terms of the potential correspond to the effective F-term potential for the
volume modulus in the large volume models. The structure of these terms generates a minimum at large values
of $\Phi$, $\Phi \sim \epsilon^{-2/3}$. The definition of $\Phi$ as $\sqrt{\frac{2}{3}} \log \mc{V}$
implies that this is equivalent to the existence of a
minimum at exponentially large volumes. As in string theory
the gravitino mass is given by $m_{3/2} = M_P (W_0/ \mc{V})$, where $W_0$ is the constant (flux) superpotential,
hierarchically large volumes corresponds to hierarchically low gravitino mass.
%If $\epsilon = 0.008$ the location of the minimum is at $\mc{V} \sim 10^{15}$, as appropriate for generating TeV supersymmetry.
The final term, depending on $\delta$, is the uplift term
needed to lift the minimum of the potential to Minkowski space.

The second and third terms, depending on $C$ and $D$, are only important at small volumes.
These terms are not derived from string theory but are included in a phenomenological fashion in order
to generate an inflection point, and thus inflation, at small volume.
Besides the large volume minimum, occurring for large values of
$\Phi$ for which the $C$ and $D$ terms are
irrelevant, by tuning $C$ and $D$\footnote{Notice that in full string compactifications,
corrections to the K\"ahler
  potential such as may generate the parameters $C,D$ will be functions of both fluxes and complex
  structure moduli and so would be expected to take on many
  possible values in the landscape.}
this sum of exponentials can have a non-monotonic shape
for relatively
small values of the field $\Phi$.

\begin{figure}[h!]
\centering{\includegraphics[height=5cm]{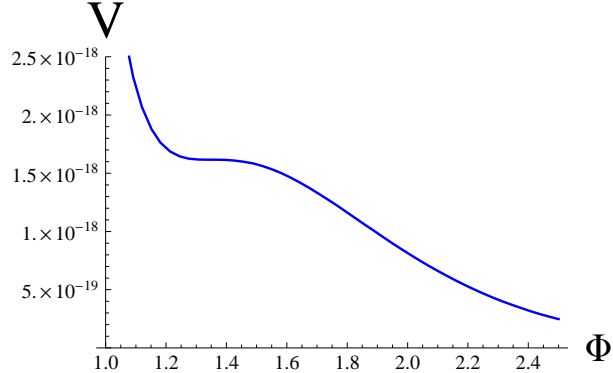}}
 \caption{The potential at small $\Phi$. Inflation occurs near the inflection point at $\Phi \sim 1.3$, in Planck units.} \label{potbeginning}
\end{figure}

We will consider the following set of parameters, which are determined by the requirement of obtaining inflation satisfying the constraint which follow from the existing observational data: $V_{0} = 1.45\times 10^{{-14}}$, $\epsilon = 0.013$, $C = -3$, $D = 2.3045$, $\delta = 0.06155\times 10^{{-10}}$. The shape of the potential for this parameters at small $\Phi$ is shown in Fig. \ref{potbeginning}.  As we see, it has an inflection point at $\Phi \sim 1.3$, where inflation takes place. The Hubble constant at that time is about $10^{-9} M_P \sim 10^{9} $ GeV.

The potentials of this type have been used by many authors, see e.g. \cite{Holman:1984yj,MSSM,Baumann:2007np,delicate,Krause:2007jk,Panda:2007ie,Itzhaki:2007nk,Linde:2007jn,Underwood:2008dh}. The typical feature of this class of models is that by tuning their parameters, the potential can always be represented as a sum of a linear term and a cubic term in a vicinity of the inflection point,
\be
V = V(\Phi_{0})\, \left (1 + \lambda_{1} (\Phi - \Phi_{0}) + {\lambda_{3}\over 3 }(\Phi - \Phi_{0})^{3}+...\right).
\ee
Here $\Phi_{0}$ is the position of the inflection point. (In our case, $\Phi_{0} \approx 1.3$, in Planck units.) The numerical values of $\lambda_{i}$ are determined by fine-tuning of the values of the parameters in our full expression for $V$. By making the linear term as small as possible, one can maximize the exponential growth of the universe during inflation, which may serve as a possible justification of the fine-tuning of the potential required for a long stage of inflation \cite{Linde:2007jn}.

In the case where inflation is dominated by the cubic term, inflation starting at the inflection point is eternal, the spectral index $n_{s} \approx 0.93$, and the amplitude of perturbations of metric produced during inflation satisfies the COBE-WMAP normalization for
\be
{V(\Phi_{0})} \approx  3\cdot 10^{-14}\, \lambda_3^{-2}
\ee
in units of Planck density  \cite{Linde:2007jn}. We calculated $V(\Phi_{0})$ and $\lambda_{i}$ and tuned the parameters of our potential to satisfy this constraint and make $\lambda_{1}$ vanishingly small.
(If one reduces the degree of fine-tuning and considers a theory with a non-vanishing negative $\lambda_{1}$, one can significantly increase $n_{s}$, but it will simultaneously decrease the degree of expansion of the universe during inflation \cite{Linde:2007jn}.)

\begin{figure}[h!]
\centering{\includegraphics[height=5cm]{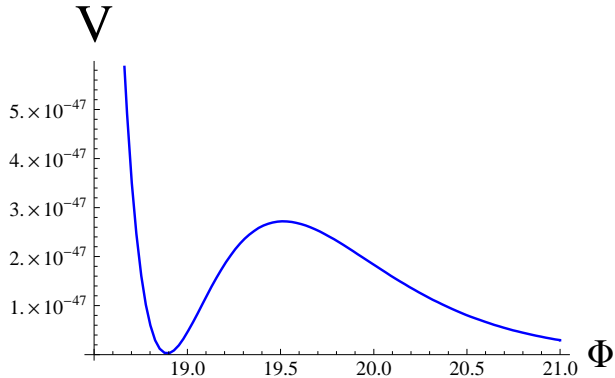}}
 \caption{The potential at large $\Phi$. Vacuum state corresponds to the minimum at  $\Phi \sim 19$.} \label{potend}
\end{figure}

After inflation, the field enters a runaway period until it is captured by the minimum at $\Phi \sim 19$, see Fig. \ref{potend}.
The parameters of our model were tuned to make the vacuum energy nearly zero in the minimum, and to have the gravitino mass there in the TeV range, $m_{3/2} = O(1) \rm{TeV}$. This last statement assumes that the gravitino mass is determined entirely by the volume, taking
$W_0 \sim 1$ as usually done in the large volume models of moduli stabilisation.

Let us study the cosmological dynamics of rolling moduli in the potential
(\ref{1modpot}).
The $C$ and $D$ terms are only important at small $\Phi$ and are highly suppressed for large $\Phi$.
Conversely, the $\delta$ and $\epsilon$ terms only become important at large
$\Phi$. These generate the minimum
at exponentially large volumes and uplift it to Minkowski space.
However, for smaller values of the volume
these terms contribute highly subleading corrections to the potential.
The upshot is that for a very large range of the volume
the dominant term in the potential (\ref{1modpot}) is simply
\be
\label{pureexp}
V = V_0 e^{- \sqrt{\frac{27}{2}} \Phi},
\ee
with $\Phi$ canonically normalized. In this regime the cosmological dynamics reduces to that
of a pure exponential potential.

Evolution along runaway directions of this type is associated with the overshoot problem \cite{bs} and
it is necessary that the fields are able to locate the global minimum of the potential. In the presence of any additional
sources (such as radiation) exponential potentials have attractor
solutions in which the different components of energy track the total energy as a constant fraction.
This attractor solution can guide the fields to the global minimum without overshooting.
In order to avoid decompactification and the overshooting problem, it is necessary that
a small component of radiation be present in the evolution equations.
We study the evolution with radiation present immediately at the end of
inflation.\footnote{A deficiency of the 1-modulus potential is that there is no way to
generate this radiation, as any primordial radiation is diluted during inflation, and thus
it has to be introduced by hand. With more complicated models, for example the 2-modulus case studied in the next
section, there are ways to generate radiation at the end of inflation.}

The evolution of fields in pure exponential potentials (\ref{pureexp}) has been studied for a long time.
Naively $\Phi$ will rapidly roll down the exponential slope into a kination phase.
However in the presence of a background fluid, such as radiation, this does not
occur
\cite{grqc9711068,astroph9711102,hepph0506045,Wetterich88,overshoot,hepth0408160,init,
hepth0505098}.
Since kinetic energy density falls faster than density  of radiation,
$$
\rho_{rad} \sim \frac{1}{a^4}, \qquad \rho_{KE} \sim \frac{1}{a^6},
$$
the radiation comes to dominate the energy density, contributing additional Hubble friction to the field dynamics.
The late-time attractor is a scaling solution in which
the different components (radiation, potential energy, kinetic energy) constitute fixed fractions of the overall
energy density.

Following \cite{grqc9711068, hepph0506045}, the equations of motion are conveniently formulated in terms of the variables $x$ and $y$ defined by
$$
x = \left( \frac{\dot{\Phi}}{\sqrt{2}} \right) \frac{1}{\sqrt{3} H}, \qquad y = \frac{\sqrt{V}}{\sqrt{3} H}.
$$
$x^2$ and $y^2$ equal $\Omega_{\Phi, kin}$ and $\Omega_{\Phi, pot}$ respectively.
It is also convenient to use $N = \ln a$ as the time variable. The equations of motion are
\bea
x'(N) & = & -3x - \frac{V'(\Phi)}{V} \sqrt{\frac{3}{2}} y^2 + \frac{3}{2} x \left( 2 x^2 + \gamma ( 1 - x^2 - y^2) \right), \\
y'(N) & = & \frac{V'(\Phi)}{V} \sqrt{\frac{3}{2}} x y + \frac{3}{2}y \left( 2 x^2 + \gamma (1 - x^2 - y^2 ) \right), \\
H'(N) & = & - \frac{3 H}{2} \left( 2 x^2 + \gamma ( 1 - x^2 - y^2) \right), \\
\Phi'(N) & = & \sqrt{6} x.
\eea
$\gamma = 4/3$ for radiation and $1$ for matter.
For an exponential
potential $V = V_0 \exp{ (- \lambda \Phi)}$ with $\lambda^2 > 6$, there exists a scaling solution
given by \cite{Wetterich88,astroph9711102,grqc9711068}
$$
x^2 = \frac{3}{2} \frac{\gamma^2}{\lambda^2}, \qquad y^2 = \frac{3( 2- \gamma) \gamma}{2 \lambda^2},
\qquad \Omega_{\Phi} = \frac{3 \gamma}{\lambda^2}.
$$
For the case at hand of $\lambda = \sqrt{\frac{27}{2}}$ and $\gamma = \frac{4}{3}$, the attractor solution is
\be
\label{attractorvalues}
\Omega_{\gamma} = \frac{19}{27}, \qquad \Omega_{kin, \Phi} = x^2 = \frac{16}{81}, \qquad \Omega_{V} = y^2 = \frac{8}{81}.
\ee
In the presence of any initial radiation or matter, this solution is the late-time attractor.

Considering the full potential (\ref{1modpot}), the attractor solution will exist so long as the pure exponential
provides a good approximation to the potential. At small $\Phi$, this approximation will hold
soon after inflation has ended, once the $C$ and $D$ terms become negligible.
Inflation occurs at $V \sim 10^{-18}$, corresponding to $\Phi \sim 1.3$.
At large $\Phi$, the attractor solution disappears as $\Phi$ approaches the susy breaking vacuum
and additional terms become important. We want the vacuum to satisfy
$m_{3/2} \sim 1 \hbox{TeV}$, and as the characteristic scale of the large-volume potential is
is $V \sim m_{3/2}^3  \sim 10^{-47}$ (assuming $W_0 \sim \mc{O}(1)$), this is attained when $\Phi \sim 19$, see Fig. \ref{potend}.

The regime in which the potential is well described by (\ref{pureexp}) and the attractor solution is valid is therefore
\be
\label{attractorvalidity}
1.3 \lesssim \Phi \lesssim 19,
\ee
a range of almost twenty Planckian distances. We note that despite the substantially trans-Planckian
field range, the potential is under good control: the
high-scale theory is understood and the potential is simply a decompactification potential.

\begin{figure}[h!]
\centering{\includegraphics[height=5cm]{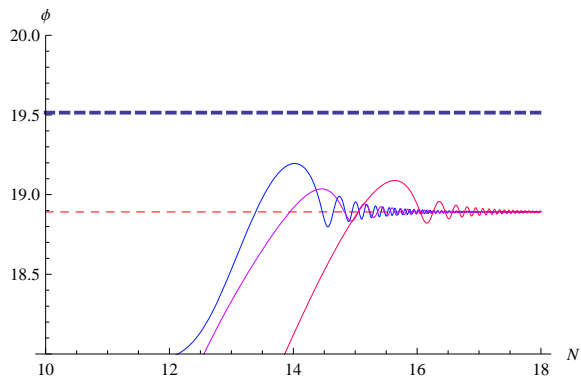}}
\caption{The evolution of the radiation-background attractor solution as it approaches the minimum. The solid dashed horizontal
line shows the location of the barrier to decompactification, and the narrow horizontal line the location of the true minimum.
The attractor solution settles at the minimum and does not overshoot. The different paths correspond to different initial conditions. $N = \ln a$ is
the time variable.}
\label{fig1}
\end{figure}

The physical minimum exists at $\Phi \sim 19$, shortly followed by a local maximum representing the
barrier to decompactification. In this regime the attractor solution is no longer present.
It is necessary that the fields fall into the minimum rather than passing over the barrier and running away to decompactification.
Whether this occurs or not depends on whether or not the field $\Phi$ has located the attractor solution while in the range
(\ref{attractorvalidity}). If the attractor solution has been
found, overshooting does not occur and $\Phi$ settles into its minimum. This is illustrated in figure \ref{fig1}, which shows the
behavior of the attractor solution as it approaches the minimum.

\begin{figure}
\twographs{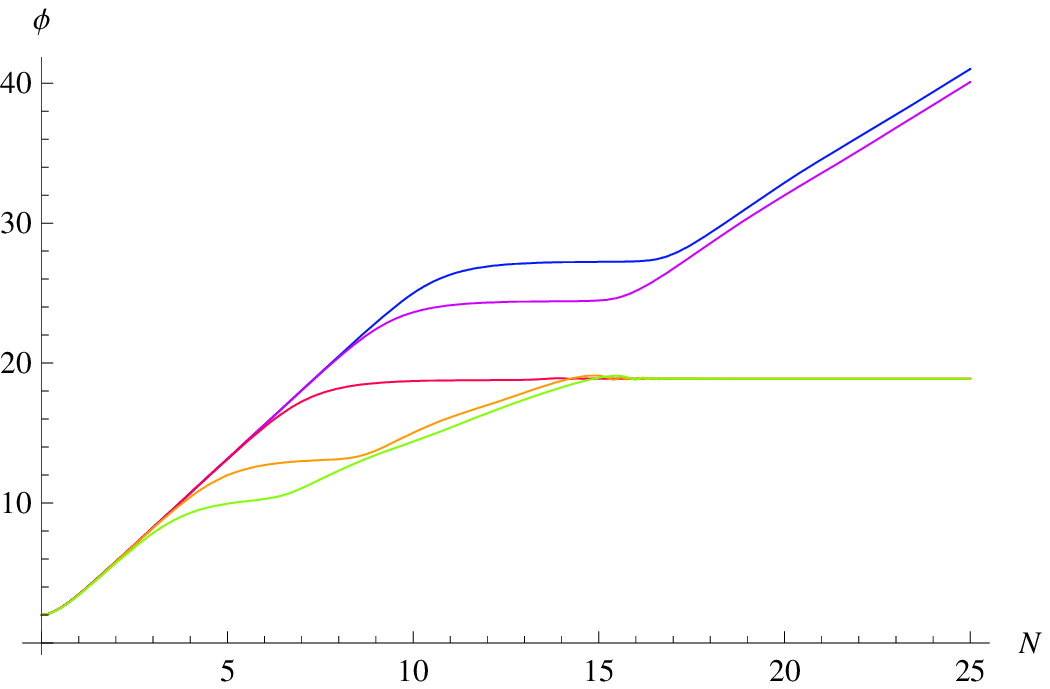}{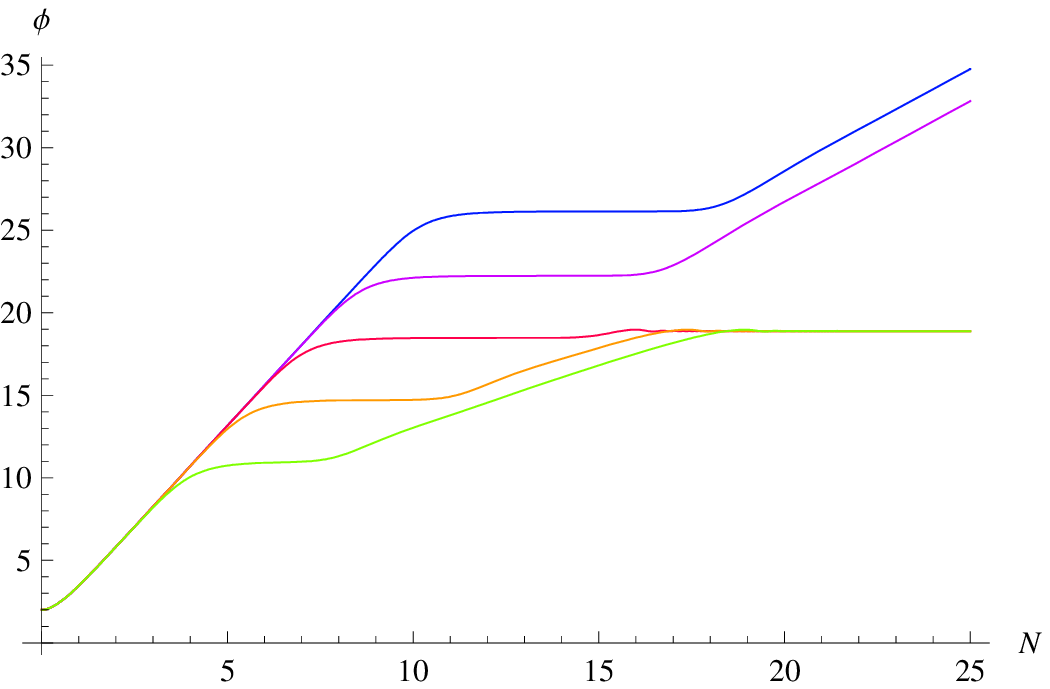}
\caption{The evolution of the field with time for radiation (a) and matter (b) backgrounds. In both cases
the field evolution is taken to start at $\Phi_{init}=2$ with $\dot{\Phi}_{init} = 0$. In figure (a), the field evolution is shown for initial radiation
densities $\Omega_{\gamma} = 10^{-7}, 10^{-6}, 10^{-4}, 10^{-2}, 10^{-1}$. For $\Omega_{\gamma} = 10^{-7}$ or $10^{-6}$
the field overshoots the minimum, whereas for the other values of $\Omega_{\gamma}$ the field finds the tracking attractor solution and settles in the minimum.
Figure (b) is for a matter background, with initial values of
$\Omega_{m} = 10^{-11}, 10^{-9}, 10^{-7}, 10^{-5}, 10^{-3}$. For $\Omega_{m} = 10^{-11}$ or $10^{-9}$ the field overshoots the minimum,
whereas for all the other values of $\Omega_{m}$ the field finds the tracking attractor solution and settles in the minimum.}
\label{taubsevolution}
\end{figure}

Whether or not the fields locate the attractor solution depends on the initial conditions and in particular
the initial amount of radiation present. This is illustrated in figure
\ref{taubsevolution}. The initial conditions were chosen to be $\Phi_{init} = 2 \approx \Phi_{0} , \dot{\Phi}=0$
with varying initial values of $\Omega_{\gamma}$. As expected, the larger the initial value of
$\Omega_{\gamma}$ the more rapidly the attractor solution is found. As long as the attractor solution is found before the
field reaches the decompactification barrier, overshooting does not occur. It is seen that even very small initial
values of $\Omega_{\gamma}$ are sufficient
to avoid overshooting, with even $\Omega_{\gamma,init} = 10^{-4}$ being sufficient to prevent overshooting.
This can be compared with the results
of \cite{hepth0408160, hepph0506045} (see in particular figure 3 of \cite{hepph0506045}), where a similar question was studied
in the context of a KKLT model,
and to avoid overshooting $\Omega_{\gamma} > 0.5$ was required across the whole range of parameter space.

In the case studied here it
is much easier to avoid overshooting, which can be achieved with $\Omega_{\gamma, init} \ll 1$.
It is easy to understand why this model is more efficient at avoiding
overshooting than the models studied in \cite{kklt, hepph0506045}.
If the runaway direction comes from gaugino condensation in either the dilaton or volume directions, then the canonically
normalized potential is a double exponential $e^{-e^{\Phi}}$ (as the potential is $\sim e^{-T}$ and the canonically
normalized field is $\Phi \sim \ln T$). This is much steeper than
the single exponential ($e^{-\Phi}$) present here, and so it is much harder to avoid overshooting.

The above one-field model has illustrated the required form of the potential, with a region suitable for inflation
at small volume, a minimum at very large values of the volume, and a region in-between where the potential is described by
a runaway. In order to locate the attractor solution that will guide the field to the global minimum, it is necessary that a small
amount of matter or radiation can be generated as runaway starts. An attractive feature is that the quantity of radiation required for the fields
not to overshoot is very small ($\Omega_{\gamma} \sim 10^{-4}$). However, any primordial radiation will be diluted away during inflation.
It is therefore necessary to extend the 1-modulus model to describe
(partial) reheating and the generation of radiation at the end of inflation, as this is necessary
to act as a source for the attractor solution.

\section{Two-Field Inflation}

We now consider a 2-field model which will justify the form of the potential used in the discussion of the 1-field model.
The model we use is the following $\mc{N}=1$ supergravity theory
\bea
\mc{V} & = & \frac{1}{9 \sqrt{2}} \left( \tau_b^{3/2} - \tau_s^{3/2} \right) \nonumber \\
K & = & -2 \ln \left( \mc{V} + \xi + \frac{C}{\mc{V}^{1/3}} + \frac{D}{\mc{V}^{2/3}} \right), \nonumber \\
W & = & W_0 + Ae^{-a_s T_s} \nonumber \\
V & = & e^{K} \left( K^{i \bar{j}} D_i W D_{\bar{j}} W - 3 \vert W \vert^2 \right).
\label{themodel}
\eea
Except for the parameters $C$ and $D$, this is the supergravity theory that describes the large-volume models for
compactifications on the Calabi-Yau manifold $\mbb{P}^4_{[1,1,1,6,9]}$ \cite{hepth0502058}.
$\tau_b$ and $\tau_s$ are K\"ahler moduli: $\tau_b$ controls the size of the Calabi-Yau, whereas $\tau_s$ is the size of
a small blow-up `hole' in the Calabi-Yau.

Although all input parameters are of order unity, this theory admits a vacuum at exponentially large
values of the volume with hierarchically small supersymmetry breaking \cite{hepth0502058}.
In (\ref{themodel}) additional corrections parametrized by $C$ and $D$ have been included. These are
motivated by the existence of higher $\alpha'$ corrections to the K\"ahler potential, which will correct the K\"ahler potential
at higher orders in the inverse volume expansion.
As discussed above, these are not derived from string theory but are included in a phenomenological fashion
to ensure an inflection point in the volume direction giving inflation at small volumes.
In this respect the detailed form of these terms
is not important - any other terms producing similar physics are
equally acceptable.

We first explain the relation between the model of (\ref{themodel}) and the 1-modulus example discussed in the previous section.
We shall work at large volume, in the regime of parameter space that applies during the runaway period, and start by
dropping the $C$ and $D$ terms of equations (\ref{themodel}), which rapidly become irrelevant at large values of the volume.
We also drop terms suppressed by higher orders in volume.
It is in this region that the 1-modulus model is a precise limiting case of the 2-modulus potential.
 The supergravity
scalar potential for the large-volume model of (\ref{themodel})
is
\be
\label{pot2}
V = \frac{8 \sqrt{\tau_s} (a_s A_s)^2 e^{-2 a_s \tau_s}}{3 \mc{V}} - \frac{4 W_0 a_s A_s \tau_s e^{-a_s \tau_s}}{\mc{V}^2}
+ \frac{\xi W_0^2}{2\mc{V}^3} + \frac{\delta}{\mc{V}^2}
\ee
We have included an uplift term (parametrized by $\delta$) to ensure the physical minimum has vanishing cosmological constant.

There are two fields, $\tau_s$ and $\tau_b$.
These two fields have very different characteristic mass scales. Working in the vicinity of
$e^{-a_s \tau_s} \sim \mc{V}^{-1}$, we have (with $M_P = 1$) \cite{hepth0502058}
\be
m_{\tau_s}^2 \sim K^{s \bar{s}} \frac{\partial^2 V}{\partial \tau_s^2} \sim \frac{1}{\mc{V}^2},
\qquad m_{\tau_b}^2 \sim K^{b \bar{b}} \frac{\partial^2 V}{\partial \tau_b^2} \sim \frac{1}{\mc{V}^3}.
\ee
As the scale of the potential is $V \sim H^2 \sim \frac{1}{\mc{V}^3}$,
it follows that $m_{\tau_s} \gg m_{\tau_b} \sim H$. We stress that these relations hold in the large-volume
regime of the potential, which will correspond to the runaway period after inflation has ended. In this regime
it is therefore possible to integrate out the heavy $\tau_s$ field
to generate a single-field potential for the volume modulus $\tau_b$.
This approximation will work with increasing accuracy as the volume increases, as the parametric separation between the masses of
$\tau_s$ and $\tau_b$ becomes increasingly large.
Now,
\bea
\label{tauseq1}
\frac{\partial V}{\partial \tau_s} & = & 0 \rightarrow (a_s A_s) e^{-a_s \tau_s} = \frac{W_0 \sqrt{\tau_s}}{2 \mc{V}} \left( 1 +
\mc{O}\left(\frac{1}{a_s \tau_s} \right) \right) \nonumber \\
& = & \frac{W_0 (\ln \mc{V})^{\half}}{2 \sqrt{a_s} \mc{V}} \left(1 + \mc{O}\left(\frac{1}{\ln \mc{V}}\right)\right).
\eea
We can use equation (\ref{tauseq1}) to integrate out $\tau_s$. Up to terms subleading in $\ln \mc{V}$, this generates a potential
$$
V = - \frac{4 W_0^2 (\ln \mc{V})^{3/2}}{3 a_s^{3/2} \mc{V}^3} + \frac{\xi W_0^2}{2\mc{V}^3} + \frac{\delta}{\mc{V}^2}.
$$
Relating $\mc{V} \sim \frac{1}{9\sqrt{2}} \tau_b^{3/2}$ to $\Phi$ using $\Phi=\sqrt{3/2}\, \log (\tau_b)$ we see that this
potential indeed takes the form of the 1-modulus potential (\ref{1modpot}).
The $C$ and $D$ terms generate higher corrections of order $\mc{V}^{-10/3}$ and $\mc{V}^{-11/3}$, as appropriate to match onto
(\ref{1modpot}). We emphasise that at smaller values of the volume,
and in particular during inflation, the above procedure of integration out is not valid. In this regime the two potentials
are qualitatively similar - in both cases there is an inflection point in the volume direction - but it is not the case that
the one-modulus model arises as a strict limit of the 2-modulus model.

Having explained the relation of the 2-modulus model to the simpler 1-modulus case studied in the previous section,
we now turn to the full two-field system described by (\ref{themodel}).
We use the following numerical parameters
$$
\xi = 9, \qquad C = -173.405, \qquad D = 1200, \qquad W_0 = -0.1, \qquad A = 1, \qquad a_s = \frac{2 \pi}{4}.
$$
The choice of numerical parameters is such that a minimum exists at exponentially large volumes with a TeV gravitino mass, while the
energy scale during inflation is $V \sim 10^{-17} M_P^4$. As this model has various phenomenological difficulties,
to be explained further below, we shall not
attempt a detailed comparison with observation.
The parameters $C$ and $D$ are chosen such that
at smaller values of the volume there exists
an inflection point in the volume direction. The large flux numbers present in flux compactifications
(typically $\int G_3 \wedge \bar{G}_3 \sim \chi(M) \sim 500$) make the relatively large values of $C$ and $D$ not unreasonable.
We choose the initial conditions so
that fields start near the inflection point, where inflation will occur,
ending with the volume rolling away from the inflection point towards
a decompactification limit.
\begin{figure}[h!]
\centering{\includegraphics[height=5cm]{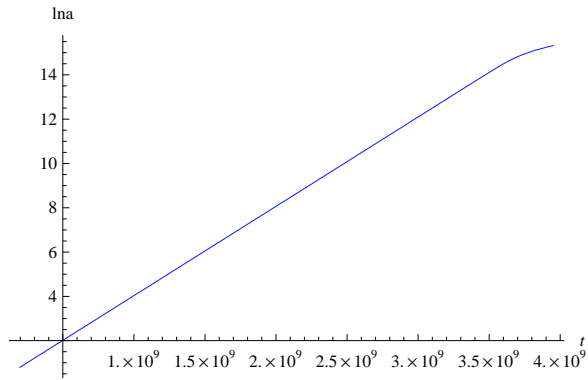}}
\caption{The evolution of the scale factor with time.}
\label{scalefactorplot}
\end{figure}

In addition to the volume direction, there is also the $\tau_s$ direction.
As the corrections to the K\"ahler potential of (\ref{themodel}) depend only on the volume, they do not induce a potential for the
$\tau_s$ direction at constant volume. This direction is lifted only
by non-perturbative superpotential terms and so is flat at large $\tau_s$.
This is equivalent to the method used to generate a flat inflationary direction in K\"ahler moduli
inflation \cite{hepth0509012, roulette}.
In the vicinity of an inflection point for
the volume direction, and with the $\tau_s$ direction lifted only by non-perturbative effects, the resulting potential is therefore
flat and suitable for slow-roll inflation.\footnote{There may be additional loop corrections in $\tau_s$ that would lift this flatness,
but these are beyond the scope of this paper.}

We choose initial conditions in which the fields have no initial velocities and have initial values
$$\tau_{b,init} = 4440, \, \, \tau_{s,init} = 23.56962555$$
\begin{figure}
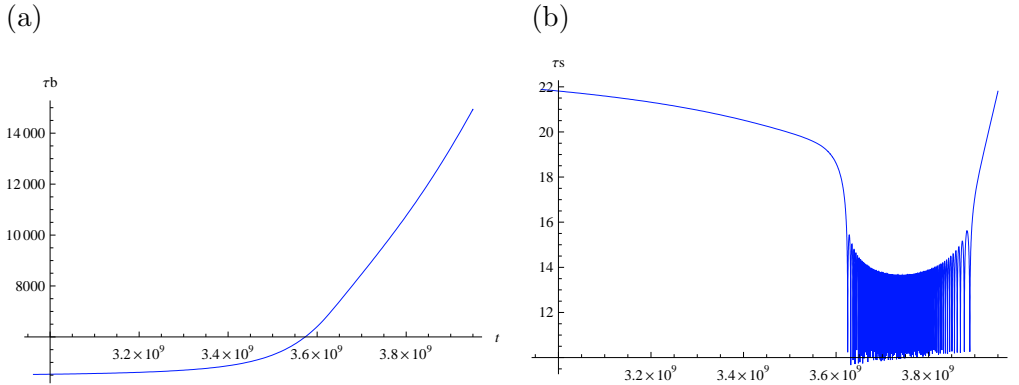

\twographs{ {taubevolution}}{{tausevolution}}
\caption{The evolution of the $\tau_b$ and $\tau_s$ fields with time. The end of inflation is marked by runaway in the $\tau_b$ direction
and oscillations in the $\tau_s$ direction.}
\label{taubsevolution1}
\end{figure}
Figure \ref{scalefactorplot}
shows the evolution of the scale factor and figure \ref{taubsevolution1} shows the evolution of the fields with
time.
For numerical reasons when performing the numerical integration
we have restricted to the last ten efolds, but there is no conceptual difficulty in extending this to sixty efolds.
In figure \ref{contourplot} we show the form of the field evolution.
Inflation occurs early in the
field trajectory, followed by a period of oscillations.
The initial conditions used are chosen to ensure that as inflation ends there is runaway in the volume
direction (the $\tau_b$ direction) and oscillations in the $\tau_s$ direction.
The presence of oscillations ensures that some quantity of post-inflationary radiation will be generated by
the decays of the oscillating field.
The large fine-tuning of initial conditions is primarily due to the need to ensure oscillations, and thus radiation,
are present at the end of inflation, as it is the radiation that allows the attractor solution to exist and be located.
 In the appendix we show that the tracker solution present in the 1-modulus case is also
present in the 2-modulus case. Numerically, we can confirm that this is also an attractor.
This radiation can act as a seed allowing the attractor
solution to be located. It is the tracker solution that will guide the fields to the global minimum.

\begin{figure}[h!]
\centering{\includegraphics[width=13cm, height=8cm]{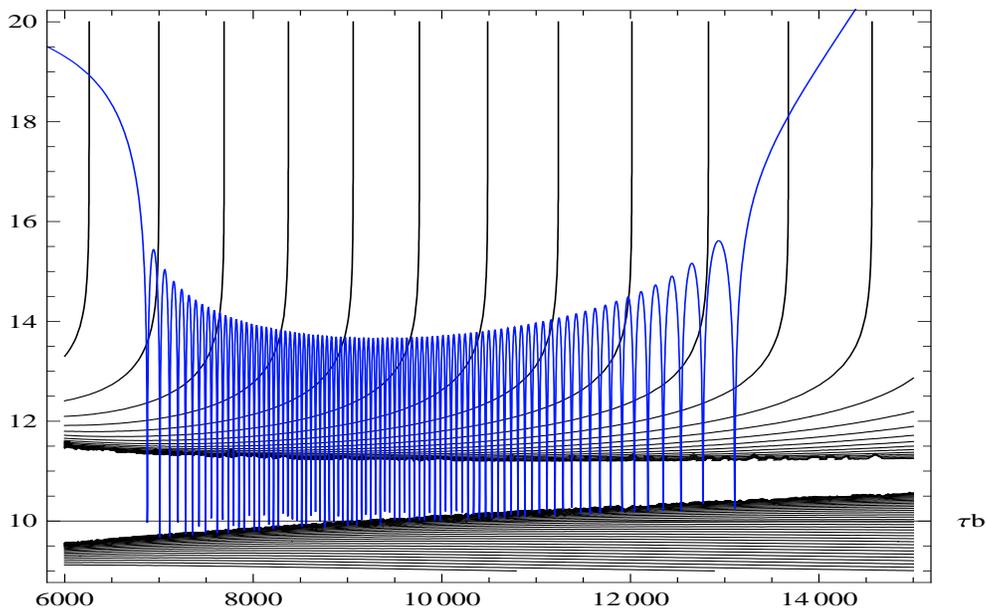}}
\caption{A plot of the field evolution, showing the inflationary era followed by oscillations in the $\tau_s$ direction.}
\label{contourplot}
\end{figure}

In figures \ref{taubsevolution1} and \ref{contourplot}
we see that with a purely classical evolution the magnitude of oscillations of $\tau_s$ grows with time.  If the magnitude of
oscillations is not sufficiently reduced by particle decays, then in fact the oscillatory $\tau_s$ field escapes from its
stabilized location. In this case there is no way to find the global minimum of the scalar potential and the solution will
decompactify.
Particle decays are intrinsically quantum events and are not explicitly included in the numerical evolution
shown in figures \ref{taubsevolution}. If we assume a brane wrapping the cycle $T_s$, then the $\tau_s$ direction couples
to radiation through the Lagrangian terms
\bea
\mc{L} & = &  \lambda \tau_s F_{\mu \nu} F^{\mu \nu} + K_{s \bar{s}} \partial_\mu \tau_s \partial^\mu \tau_s
+ V(\tau_s) %\\
%& = &  \tau_s F_{\mu \nu} F^{\mu \nu} + K_{s \bar{s}} \partial_\mu \tau_s \partial^\mu \tau_s
%+ V_0 + \half \frac{\partial^2 V}{\partial \tau_s^2} + \ldots
\eea
From this the decay width for $\tau_s \to \gamma \gamma$ can be
computed
\cite{cq},
\be
\Gamma \sim \frac{\lambda^2 K^{s \bar{s}} m_{\tau_s}^3}{M_P^2}.
\ee
Evaluating this for the model of \ref{themodel}, we have
$$
K^{s \bar{s}} \sim \mc{V}, \qquad m_{\tau_s} \sim \frac{(\ln \mc{V}) M_P}{\mc{V}}, \qquad H \sim \frac{M_P}{\mc{V}^{3/2}}
$$
 The most important relation is the relative size of the decay width $\Gamma$ and the Hubble scale $H$. The volume scaling of this
is given by
$$
\frac{\Gamma_{\tau_s \to \gamma \gamma}}{H} =  \frac{\lambda^2 (\ln \mc{V})^3}{\sqrt{\mc{V}}}.
$$
The question of whether the decays reduce the oscillation amplitude sufficiently rapidly is determined by
the model-dependent details of the
coupling of $\tau_s$ to radiation and on the initial conditions. If necessary it is always possible to tune the initial conditions
to ensure there are sufficient oscillations of the $\tau_s$ field in order to generate sufficient decays to radiation.

The most important point that this model illustrates is that inflation ends with runaway in the volume direction
together with a small quantity of radiation. The radiation will serve as a seed for the attractor solution
that will guide the fields into the global minimum.
In the specific case of the model above the radiation is
provided by decays of $\tau_s$ oscillations.

\section{Conclusions and Challenges}

Let us discuss what has been done in this paper and what remains to be
done. This paper has proposed a mechanism in which high-scale inflation and low scale
supersymmetry can coexist. The model discussed explicitly is by no means problem-free and should be regarded primarily as
an illustration of the main features of the proposed mechanism. These involve a runaway at the end of
inflation, with the true vacuum far away from that applicable during inflation.

Let us discuss some open questions for this
explicit model. First, volume modulus inflation was obtained by fine tuning the
parameters $C$ and $D$ to generate an inflection point. The amount of fine tuning is at a similar order to that
required in models of brane-antibrane inflation but is not
explicit as it relies on higher order quantum corrections to
the K\"ahler potential that are not controlled. The use of the volume
modulus for the inflaton was only for simplicity and it is an
interesting challenge to improve the implementation of
inflation. One interesting generalization would be to a hybrid
model of inflation in which the decompactification direction that generates the runaway
plays the role of the tachyonic field that ends inflation.

A second issue relates to the generation of radiation and isocurvature perturbations. In the
inflationary regime, we had the inflaton (volume) modulus light
to drive inflation but also required the second (small) modulus to be light.
We also needed a large fine-tuning of the initial conditions.
This was to ensure that the small modulus would oscillate at the end of inflation and generate through its decays
the radiation necessary for the attractor solution.
Indeed, if the second field is very heavy, it always follows the time-dependent position of its potential determined by the slowly-changing inflaton. In this case it will not experience oscillations after the end of inflationary stage, and so it would not be possible
to generate the radiation required for the attractor solution.

The lightness of this modulus is a source of isocurvature perturbations that are highly constrained observationally. For the purposes of this paper, which aims simply at combining high scale inflation with TeV supersymmetry, this is less of an issue.
To make our model fully realistic, one may consider the second
field to be a curvaton and later on having the isocurvature
perturbations converted to adiabatic perturbations \cite{curvaton1}.  Another possibility to use the second light field for generation of adiabatic perturbations is discussed in \cite{curvaton2}. In both cases, inflation may produce perturbations of metric with a measurable degree of non-gaussianity. This can be a potential advantage of our model, having in mind the recent controversy with respect to the possible non-gaussianity in the WMAP data  \cite{Yadav:2007yy,Hikage:2008gy,McEwen:2008kv,Komatsu:2008hk}.

%Furthermore,
On the other hand,
the fact that both moduli are light in the inflation
regime could be modified by additional quantum corrections to the K\"ahler
potential. Such corrections are potentially a problem for K\"ahler moduli
inflation as proposed in \cite{hepth0509012}. In that case the flatness
of the potential for the small modulus is natural up to and including the leading
$\alpha'$ corrections to the K\"ahler potential. String loop
corrections \cite{bhk} not included in \cite{hepth0509012} have the
potential to lift that flatness. A correction of the
form $\delta K \sim 1/({\cal V} \tau_s)$ would give a dominant contribution to
the mass of $\tau_s$, resurrecting the $\eta$ problem for K\"ahler
moduli inflation but avoiding the second light direction in our case.

All the above issues refer specifically to the particular inflationary model described in this paper.
In general, for the proposed mechanism to occur the
principal requirements on the inflationary model are only that at the end of inflation
there is both a small amount of radiation present and runaway in the volume direction. In this respect
the overall scenario is not tied to the particular model used in this paper.
The requirement of radiation at the end of inflation is suggestive of models of brane inflation,
where the brane annihilation that ends inflation will generate large quantities of radiation.

Finally, as generally holds for models of cosmology with
TeV-scale supersymmetry, this scenario still suffers from the cosmological moduli
problem \cite{cmp}. This was recently re-analyzed in \cite{cq} for
the large volume class of models. While the small moduli decay very rapidly,
the volume modulus has an MeV-scale mass
and gravitational strength interactions. It naturally decays very late in the
history of the universe ruining successful nucleosynthesis. An often discussed solution of this problem is to have a late period of inflation to dilute the modulus, such as thermal inflation \cite{thermal}. Even
though an explicit realization of thermal inflation in our model is
not at hand, the attractor solution automatically contains a high fraction of radiation at the end of the
runaway period, which may provide the right initial conditions for thermal
inflation.

Thus we see that in principle there exist various possibilities to avoid the bound $H\lesssim m_{3/2}$ in string cosmology. One of these possibilities is to study inflation in  KL models \cite{kl}, or in the model proposed recently in \cite{Badziak:2008yg}. Yet another possibility is to construct a cosmological scenario along the lines outlined in this paper. However, none of these possibilities is easy and all of them require fine-tuning. This may just mean that the constraint  $H\lesssim m_{3/2}$ is quite robust and hard to avoid, so one should consider inflationary models where inflation occurs on a very small energy scale.

It is worth emphasizing the fact that low-energy
supersymmetry, especially within string constructions, tends to put very
strong constraints on viable cosmological scenarios. The constraint $H\lesssim m_{3/2}$ is just one of them. Other constraints include the cosmological problems associated with light gravitino and moduli fields \cite{gravitinos,Coughlan:1983ci}. These cosmological difficulties may lead us to reconsider how much weight to give to the supersymmetric
solution of the hierarchy problem.  As
mentioned in the Introduction, one route out is to have models
of high scale supersymmetry breaking. These automatically avoid the
cosmological problems associated to the moduli fields and the gravitino
but replace the standard solution to the hierarchy problem by appeals to anthropic considerations and string theory landscape.
 On the other hand we can take a positive view on
the existence of so many cosmological constraints: they provide powerful
guidelines for the properties of realistic models and eliminate many
potential vacua that would be otherwise consistent.
Fortunately, experimental evidence will soon tell us if low-energy
supersymmetry is realized in nature in a long anticipated way.

If supersymmetry is indeed discovered at the LHC, it would falsify many proposed models of inflation in both string
and field theory. It would then be of extreme importance to understand how low-scale supersymmetry can
be consistent with the observational evidence for inflation at very high
energy scales $V_{inf} \gg 10^{11} \hbox{GeV}$, particularly if tensor perturbations were observed by either
WMAP or Planck.
In this paper we have proposed that this can occur
if inflation ends with runaway towards a decompactification direction.
The energy scales during inflation and the energy scales $\emph{in vacuo}$ can then be hierarchically different.
As long as some radiation is present or generated at the end of inflation, an attractor solution can guide
the fields into the global minimum of the potential, avoiding the cosmological overshoot problem.
We have presented a model to illustrate this
scenario, where the volume modulus was the inflaton
and the runaway direction was towards the large volume minimum.
The ultimate aim of model-building is to construct a model that is
fully realistic when confronted with both cosmological and phenomenological
data - our work is one step towards this goal.

\acknowledgments{} We thank M. Cicoli, E. Copeland, S. Sarkar, H. Tye
and especially C.P. Burgess for interesting discussions. JC is funded
by Trinity College, Cambridge and also thank the University of Texas
at Austin for hospitality while some of this work was carried out. He was supported in part by
NSF grant PHY-0455649.
FQ is partially funded by STFC and a Royal Society Wolfson merit
award and thank the Mitchell family and the organizers of the
Cambridge-Texas A\&M Cook's Branch meeting.
 RK and AL were supported in part by NSF grant PHY-0244728.  They are grateful for the hospitality and support extended to
them at  YITP, Kyoto.

\section*{A: Tracker solution for two modulus model}
\label{2fm}

In section \ref{1fm} we studied the tracker solution in the presence of radiation
for the single field potential that arises from integrating out the heavy $\tau_s$ mode.
Here we show the existence of a tracker solution in the presence of radiation for the full two-modulus model with non-canonical kinetic terms.
The tracker solution will apply after the end of inflation, namely during the runaway regime of the potential.
The model is described by
$$
\mc{V} = \frac{1}{9 \sqrt{2}} \left( \tau_b^{3/2} - \tau_s^{3/2} \right), \qquad
T_b = \tau_b + i c_b, \qquad T_s = \tau_s + i c_s.
$$
$$
K = -2 \ln \left( \mc{V} + \xi \right), \qquad W = W_0 + Ae^{-a_s T_s},
$$
$$
V = e^{K} \left( K^{i \bar{j}} D_i W D_{\bar{j}} W - 3 \vert W \vert^2 \right).
$$
Here we have used simply the potential arising for the large volume models, without the additional terms $C$ and $D$ that were
only important at small volumes.
Starting with a Lagrangian
\be
\mc{L} = \int d^4x \half \sqrt{g} R + K_{i \bar{j}} \partial_\mu \Phi^i \partial^\mu \Phi^j + V(\Phi, \bar{\Phi}),
\ee
the equations of motion for the complex fields $\Phi^i$ are
\bea
\ddot{\Phi}^i + 3 H \dot{\Phi}^i + \Gamma^{i}_{jk} \dot{\Phi}^j \dot{\Phi}^k + K^{i \bar{j}} \partial_{\bar{j}} V & = & 0, \\
\label{Friedmann}
H^2 - \frac{1}{3}\left( \mc{K}_{i \bar{j}} \dot{\Phi}^i \dot{\Phi}^j + V + \rho_{\gamma} \right) & = & 0,
\eea
with $\rho_{\gamma}$ the background (radiation) energy density and $\Gamma^{n}_{ij} = \mc{K}^{n \bar{l}} \frac{\partial \mc{K}_{j \bar{l}}}{\partial \Phi^i}$.
We fix the axionic parts of $T^b$ and $T^s$ and write the equations solely in terms of the real parts $\tau_b$ and $\tau_s$.
The equations of motion are then
\be
\label{realeqs}
\ddot{\tau}^i + 3 H \dot{\tau}^i + \Gamma^{i}_{jk} \dot{\tau}^j \dot{\tau}^k + \half K^{i \bar{j}} \partial_{\tau_j} V = 0.
\ee
It is useful to rewrite the Friedmann equation (\ref{Friedmann}) as
\be
\dot{H} = - K_{i \bar{j}} \dot{\tau}_i \dot{\tau}_j - \frac{\gamma}{2} \rho_\gamma.
\ee
In a tracker solution the kinetic, potential and background energies are all constant fractions of the total energy.
This implies $\rho_\gamma \propto H^2$, $K_{i \bar{j}} \dot{\tau}_i \dot{\tau}_j \propto H^2$, and $V(\tau) \propto H^2$, giving
\be
\dot{H} \propto -H^2,
\ee
and so in any tracker solution
\be
H = \frac{\lambda}{t}, \qquad \rho_{\gamma} \sim \frac{1}{t^2}, \qquad K_{i \bar{j}} \dot{\tau}_i \dot{\tau}_j \sim \frac{1}{t^2},
\qquad V \sim \frac{1}{t^2}.
\ee
The assumption of a radiation background, for which $\rho_{\gamma} \sim a^{-4}$, then implies $a \sim t^{\half}$ and $H = \frac{1}{2t}$.
As the magnitude of the potential is determined in terms of the compactification volume by $V \sim \mc{V}^{-3} \sim \tau_b^{-9/2}$,
we can also deduce that
$$
\mc{V} \sim t^{2/3}, \qquad \tau_b = A t^{4/9}.
$$
As at its minimum the $\tau_s$ direction has characteristic mass $m_{\tau_s} \gg H$, we anticipate that in a tracker solution
$\partial_{\tau_s} V \sim 0 \ll V$. We will show below that this assumption is self-consistent.
From equation (\ref{tauseq1}) this implies
\be
\label{tauansatz}
\tau_s \sim \ln \mc{V} \sim \frac{3}{2} \ln \tau_b = \frac{2}{3} \ln t + \rm{constant} + \ldots
\ee
We can now check that the above ansatz is consistent and indeed leads to a solution of the equations of motion.

Considering the leading terms at large volume,
the K\"ahler potential $K = - 2 \ln (\tau_b^{3/2} - \tau_s^{3/2})$ gives the following K\"ahler metric
\be
K = \left( \begin{array}{cc} \frac{3}{4 \tau_b^2} & \frac{-9 \sqrt{\tau_s}}{8 \tau_b^{5/2}} \\
\frac{-9 \sqrt{\tau_s}}{8 \tau_b^{5/2}} & \frac{3}{8 \sqrt{\tau_s} \tau_b^{3/2}} \end{array} \right), \qquad
K^{-1} = \left( \begin{array}{cc} \frac{4 \tau_b^2}{3} & 4 \tau_s \tau_b \\
4 \tau_s \tau_b & \frac{8 \sqrt{\tau_s} \tau_b^{3/2}}{3} \end{array} \right).
\ee
From this we can compute the Christoffel symbols
$$
\Gamma^{b}_{bb} = - \frac{1}{\tau_b}, \qquad \Gamma^b_{bs} = \frac{3 \sqrt{\tau_s}}{4 \tau_b^{3/2}}, \qquad
\Gamma^{b}_{ss} = - \frac{3}{4 \sqrt{\tau_s \tau_b}},
$$
$$
\Gamma^s_{bb} = \frac{3 \tau_s}{4 \tau_b^2}, \qquad \Gamma^s_{bs} = \Gamma^s_{sb} = \frac{-3}{4 \tau_b}, \qquad
\Gamma^s_{ss} = \frac{-1}{4 \tau_s}.
$$
Using the above ans\"atze we evaluate
\bea
\label{bderivs}
\ddot{\tau}_b + 3 H \dot{\tau}_b + \Gamma^b_{ij} \dot{\tau}_i \dot{\tau}_j & = & \frac{2}{9} A t^{-14/9} + (\hbox{subleading in t}), \\
\label{sderivs}
\ddot{\tau}_s + 3 H \dot{\tau}_s + \Gamma^s_{ij} \dot{\tau}_i \dot{\tau}_j & = & \frac{1}{t^2} \left( \frac{4 \tau_s}{27} + \frac{1}{9}
- \frac{1}{9 \tau_s} \right) + (\hbox{subleading in t}),
\eea
We expect any tracker solution to be valid during the regime in which the
potential is dominated by the third term (the $\alpha'^3$ correction).
During this regime, $\partial_{\tau_b} V \sim \frac{-9 V}{2 \tau_b}$.
In the tracker solution, we also expect to have $\partial_{\tau_s} V \sim 0$ (as the heavy $\tau_s$ field should be fixed at its minimum).
In this case,
\bea
\half K^{b \bar{j}} \partial_{\tau_j} V & = & \half ( K^{b \bar{b}} \partial_{\tau_b} V + K^{b \bar{s}} \partial_{\tau_s} V ) \nonumber \\
& = & \half \left( \frac{4 \tau_b^2}{3} \frac{-9 V}{2 \tau_b} + 4 \tau_s \tau_b \partial_{\tau_s} V \right) = - 3 \tau_b V = -3 A t^{4/9} V.
\eea

We have here used the assumption that $\partial_{\tau_s} V \ll V$.
Comparison with equations (\ref{realeqs}) and (\ref{bderivs}) then gives
$$
V = \frac{2}{27} \frac{1}{t^2}.
$$
The $\tau_s$ equations of motion give
\bea
\half K^{s \bar{j}} \partial_{\tau_j} V & = & \half ( K^{s \bar{b}} \partial_{\tau_b} V + K^{s \bar{s}} \partial_{\tau_s} V ) \nonumber \\
& = & -9 \tau_s V + \frac{4 \sqrt{\tau_s} \tau_b^{3/2}}{3} (\partial_{\tau_s} V)
\eea
Comparison with equation (\ref{realeqs}) and (\ref{sderivs}) then gives
\be
\label{eq1}
\frac{4}{27} \frac{\tau_s}{t^2} = 9 \tau_s V - \frac{4 \sqrt{\tau_s} \tau_b^{3/2}}{3} (\partial_{\tau_s} V)
\ee
Using $V = \frac{2}{27 t^2}$, eq. (\ref{eq1}) is satisfied so long as
\be
\label{tauseq2}
\partial_{\tau_s} V =
\frac{21 \sqrt{\tau_s} V}{4 \tau_b^{3/2}}.
\ee
Equation (\ref{tauseq2}) can be solved to determine the precise value of $\tau_s$ in the tracker solution.

As this corresponds at leading order in $t$ to our ansatz (\ref{tauansatz}), this shows that our approximations
were self-consistent.
As the characteristic scale
of $\partial_{\tau_s} V$ is $V$ itself, the requirement of (\ref{tauseq2}) that
$\partial_{\tau_s} V \sim \frac{V}{\mc{V}} \ll V$ is consistent with the interpretation that
$\tau_s$ has been integrated out.

It is interesting to analyze the origin of the kinetic energy in the tracker solution. Writing
\bea
K_{ij} \dot{\tau}_i \dot{\tau}_j & = & K_{bb} \dot{\tau}_b \dot{\tau}_b + 2K_{bs} \dot{\tau}_b \dot{\tau}_s + K_{ss} \dot{\tau}_s \dot{\tau}_s
\nonumber \\ & = & \frac{4}{27 t^2} + \mc{O}(t^{-8/3}) + \mc{O}(t^{-8/3}),
\eea
we see that the kinetic energy of the solution is dominated by the motion of the light field ($\tau_b$)
and that the contributions of the heavy field $\tau_s$ to the kinetic energy vanish at large $t$.
We can verify that the above results are consistent with the single-field attractor. We have
$$
V = \frac{2}{27 t^2}, \qquad KE = \frac{4}{27 t^2}, \qquad H = \frac{1}{2t},
$$
and so $\Omega_V = \frac{8}{81}, \Omega_{KE} = \frac{16}{81}, \Omega_{\gamma} = \frac{19}{27}$.
This reproduces the attractor values of eq (\ref{attractorvalues}) for the single-field evolution.
It is thus possible to explicitly solve the equations of motion for the 2-modulus model for the runaway regime in the presence of radiation,
finding as expected the same
tracker solution as seen when the heavy modulus is integrated out.

The tracker solution requires the presence of radiation - in the absence of radiation, the small field cannot be confined within its trough
and the moduli evolution enters a kination phase which eventually leads to decompactification.
 Furthermore, the simple existence of the tracker solution does not imply that the
moduli evolution will locate the tracker solution. To locate the tracker solution, sufficient radiation needs to be generated at the
end of inflation to attract the moduli evolution into the tracker solution. Whether sufficient radiation can be generated or not
is a dynamical question that depends on the amplitude of oscillations of the small field and on its couplings
to matter - we saw in figure \ref{contourplot} that in a purely classical evolution, with no quantum decays,
 the small field cannot be confined within its trough.

As for the 1-modulus case, the tracker solution is (numerically) also found to be an attractor solution. If the fields can
locate the attractor solution, then this will guide them through the runaway epoch and into the large-volume minimum without
overshooting.


\begin{thebibliography}{99}


\bibitem{kklt}
S.~Kachru, R.~Kallosh, A.~Linde and S.~P.~Trivedi,
``De Sitter vacua in string theory,''
  Phys.\ Rev.\  D {\bf 68} (2003) 046005
  [arXiv:hep-th/0301240].
  %%CITATION = PHRVA,D68,046005;%%

 \bibitem{hepth0502058}
  V.~Balasubramanian, P.~Berglund, J.~P.~Conlon and F.~Quevedo,
``Systematics of moduli stabilization in Calabi-Yau flux
compactifications,''
  JHEP {\bf 0503}, 007 (2005)
  [arXiv:hep-th/0502058];
  %%CITATION = JHEPA,0503,007;%%
J.~P.~Conlon, F.~Quevedo and K.~Suruliz,
``Large-volume flux compactifications: Moduli spectrum and D3/D7 soft
supersymmetry breaking,''
  JHEP {\bf 0508} (2005) 007
  [arXiv:hep-th/0505076].
  %%CITATION = JHEPA,0508,007;%%



\bibitem{kl}
R.~Kallosh and A.~Linde,
``Landscape, the scale of SUSY breaking, and inflation,"
JHEP {\bf 0412}, 004 (2004)
[arXiv:hep-th/0411011];
J.~J.~Blanco-Pillado, R.~Kallosh and A.~Linde,
``Supersymmetry and stability of flux vacua,''
  JHEP {\bf 0605}, 053 (2006)
  [arXiv:hep-th/0511042];
  R.~Kallosh and A.~Linde,
``Testing String Theory with CMB,''
  JCAP {\bf 0704} (2007) 017
  [arXiv:0704.0647 [hep-th]].

%\cite{Buchmuller:2004tz}
\bibitem{Buchmuller:2004tz}
  W.~Buchmuller, K.~Hamaguchi, O.~Lebedev and M.~Ratz,
``Maximal temperature in flux compactifications,''
  JCAP {\bf 0501}, 004 (2005)
  [arXiv:hep-th/0411109].
  %%CITATION = JCAPA,0501,004;%%

  %\cite{Covi:2008cn}
\bibitem{Covi:2008cn}
  L.~Covi, M.~Gomez-Reino, C.~Gross, J.~Louis, G.~A.~Palma and C.~A.~Scrucca,
``Constraints on modular inflation in supergravity and string theory,''
  arXiv:0805.3290 [hep-th].
  %%CITATION = ARXIV:0805.3290;%%

\bibitem{Lalak:2005hr}
  Z.~Lalak, G.~G.~Ross and S.~Sarkar,
  %``Racetrack inflation and assisted moduli stabilisation,''
  Nucl.\ Phys.\  B {\bf 766}, 1 (2007)
  [arXiv:hep-th/0503178].
  %%CITATION = NUPHA,B766,1;%%

  %\cite{Silverstein:2007ac}
\bibitem{Silverstein:2007ac}
  E.~Silverstein,
 ``Simple de Sitter Solutions,''
  arXiv:0712.1196 [hep-th].
  %%CITATION = ARXIV:0712.1196;%%

  %\cite{Silverstein:2008sg}
\bibitem{Silverstein:2008sg}
  E.~Silverstein and A.~Westphal,
``Monodromy in the CMB: Gravity Waves and String Inflation,''
  arXiv:0803.3085 [hep-th].
  %%CITATION = ARXIV:0803.3085;%%

\bibitem{hepth0509012}
  J.~P.~Conlon and F.~Quevedo,
``Kaehler moduli inflation,''
  JHEP {\bf 0601}, 146 (2006)   [arXiv:hep-th/0509012].
  %%CITATION = JHEPA,0601,146;%%

\bibitem{roulette}
  J.~R.~Bond, L.~Kofman, S.~Prokushkin and P.~M.~Vaudrevange,
``Roulette inflation with Kaehler moduli and their axions,''
  Phys.\ Rev.\  D {\bf 75},  123511 (2007)
  [arXiv:hep-th/0612197]. %%CITATION = PHRVA,D75,123511;%%
  Z.~Lalak, D.~Langlois, S.~Pokorski and K.~Turzynski,
  %``Curvature and isocurvature perturbations in two-field inflation,''
  JCAP {\bf 0707} (2007) 014
  [arXiv:0704.0212 [hep-th]].
  %%CITATION = JCAPA,0707,014;%%
  R.~Holman and J.~A.~Hutasoit,
  ``Systematics of moduli stabilization, inflationary dynamics and power
 spectrum,''
  JHEP {\bf 0608} (2006) 053
  [arXiv:hep-th/0606089].
  %%CITATION = JHEPA,0608,053;%%
  A.~Misra and P.~Shukla,
   ``Large Volume Axionic Swiss-Cheese Inflation,''
  arXiv:0712.1260 [hep-th].
  %%CITATION = ARXIV:0712.1260;%%

\bibitem{MSSM}
  R.~Allahverdi, K.~Enqvist, J.~Garcia-Bellido, A.~Jokinen and A.~Mazumdar,
``MSSM flat direction inflation: slow roll, stability, fine tuning and
reheating,''
  JCAP {\bf 0706}, 019 (2007)
  [arXiv:hep-ph/0610134];
   A.~Mazumdar,
``New developments on embedding inflation in gauge theory and particle
physics,''
  arXiv:0707.3350 [hep-ph].

      \bibitem{rosssarkar}
G.~G.~Ross and S.~Sarkar,
``Successful supersymmetric inflation,''
  Nucl.\ Phys.\  B {\bf 461} (1996) 597
  [arXiv:hep-ph/9506283].
  %%CITATION = NUPHA,B461,597;%%
G.~German, G.~G.~Ross and S.~Sarkar,
``Low-scale inflation,''
  Nucl.\ Phys.\  B {\bf 608} (2001) 423
  [arXiv:hep-ph/0103243].
  %%CITATION = NUPHA,B608,423;%%






  \bibitem{gravitinos}
P.~Fayet, ``Phenomenology Of Supersymmetry,'' Talk at the XVIIth Rencontre de Moriond, Ecole Normale Superieure preprint LPTENS 82/10 (1982);
 S.~Weinberg,
 ``Cosmological Constraints On The Scale Of Supersymmetry Breaking,''
  Phys.\ Rev.\ Lett.\  {\bf 48}, 1303 (1982);
 J.~R.~Ellis, A.~D.~Linde and D.~V.~Nanopoulos,
``Inflation Can Save The Gravitino,''
  Phys.\ Lett.\  B {\bf 118}, 59 (1982); L.~M.~Krauss,
``New Constraints On Ino Masses From Cosmology. 1. Supersymmetric Inos,''
  Nucl.\ Phys.\  B {\bf 227}, 556 (1983); M.~Y.~Khlopov and A.~D.~Linde,
  ``Is It Easy To Save The Gravitino?,''
  Phys.\ Lett.\  B {\bf 138}, 265 (1984).

    %\cite{Coughlan:1983ci}
\bibitem{Coughlan:1983ci}
  G.~D.~Coughlan, W.~Fischler, E.~W.~Kolb, S.~Raby and G.~G.~Ross,
``Cosmological Problems For The Polonyi Potential,''
  Phys.\ Lett.\  B {\bf 131}, 59 (1983);
  A.~S.~Goncharov, A.~D.~Linde and M.~I.~Vysotsky,
``Cosmological Problems For Spontaneously Broken Supergravity,''
  Phys.\ Lett.\  B {\bf 147}, 279 (1984).

\bibitem{cmp}
T.~Banks, D.~B.~Kaplan and A.~E.~Nelson,
  ``Cosmological implications of dynamical supersymmetry breaking,''
  Phys.\ Rev.\  D {\bf 49} (1994) 779
  [arXiv:hep-ph/9308292];
  %%CITATION = PHRVA,D49,779;%%
B.~de Carlos, J.~A.~Casas, F.~Quevedo and E.~Roulet,
``Model independent properties and cosmological implications of the dilaton
 and moduli sectors of 4-d strings,''
  Phys.\ Lett.\  B {\bf 318} (1993) 447
  [arXiv:hep-ph/9308325].
  %%CITATION = PHLTA,B318,447;%%




  %\cite{DeWolfe:2002nn}
\bibitem{DeWolfe:2002nn}
  O.~DeWolfe and S.~B.~Giddings,
``Scales and hierarchies in warped compactifications and brane worlds,''
  Phys.\ Rev.\  D {\bf 67}, 066008 (2003)
  [arXiv:hep-th/0208123].
  %%CITATION = PHRVA,D67,066008;%%

%\cite{Arkani-Hamed:2004fb}
\bibitem{Arkani-Hamed:2004fb}
  N.~Arkani-Hamed and S.~Dimopoulos,
``Supersymmetric unification without low energy supersymmetry and  signatures
for fine-tuning at the LHC,''
  JHEP {\bf 0506}, 073 (2005)
  [arXiv:hep-th/0405159];
  %%CITATION = JHEPA,0506,073;%%
  N.~Arkani-Hamed, S.~Dimopoulos, G.~F.~Giudice and A.~Romanino,
``Aspects of split supersymmetry,''
  Nucl.\ Phys.\  B {\bf 709}, 3 (2005)
  [arXiv:hep-ph/0409232].
  %%CITATION = NUPHA,B709,3;%%

\bibitem{myers}
A.~R.~Frey, A.~Mazumdar and R.~C.~Myers,
  ``Stringy effects during inflation and reheating,''
  Phys.\ Rev.\  D {\bf 73} (2006) 026003
  [arXiv:hep-th/0508139].
  %%CITATION = PHRVA,D73,026003;%%


  %\cite{Kachru:2003sx}
\bibitem{Kachru:2003sx}
  S.~Kachru, R.~Kallosh, A.~Linde, J.~M.~Maldacena, L.~P.~McAllister \& S.~P.~Trivedi,
``Towards inflation in string theory,''
  JCAP {\bf 0310}, 013 (2003)
  [arXiv:hep-th/0308055].

\bibitem{bcsq}
C.~P.~Burgess, J.~M.~Cline, H.~Stoica and F.~Quevedo,
  ``Inflation in realistic D-brane models,''
  JHEP {\bf 0409} (2004) 033
  [arXiv:hep-th/0403119].
  %%CITATION = JHEPA,0409,033;%%



%\cite{Baumann:2007np}
\bibitem{Baumann:2007np}
  D.~Baumann, A.~Dymarsky, I.~R.~Klebanov, L.~McAllister \& P.~J.~Steinhardt,
``A Delicate Universe,''
  Phys.\ Rev.\ Lett.\  {\bf 99}, 141601 (2007)
  [arXiv:0705.3837 [hep-th]].
  %%CITATION = PRLTA,99,141601;%%


\bibitem{delicate}
  D.~Baumann, A.~Dymarsky, I.~R.~Klebanov and L.~McAllister,
``Towards an Explicit Model of D-brane Inflation,''
  JCAP {\bf 0801}, 024 (2008)
  [arXiv:0706.0360 [hep-th]].
  %%CITATION = JCAPA,0801,024;%%

  %\cite{Krause:2007jk}
\bibitem{Krause:2007jk}
  A.~Krause \& E.~Pajer,
``Chasing Brane Inflation in String-Theory,''
  arXiv:0705.4682 [hep-th].
  %%CITATION = ARXIV:0705.4682;%%

  %\cite{Panda:2007ie}
\bibitem{Panda:2007ie}
 S.~Panda, M.~Sami and S.~Tsujikawa,
``Prospects of inflation in delicate D-brane cosmology,''
  Phys.\ Rev.\  D {\bf 76}, 103512 (2007)
  [arXiv:0707.2848 [hep-th]].  %%CITATION = ARXIV:0707.2848;%%

  %\cite{Itzhaki:2007nk}
\bibitem{Itzhaki:2007nk}
 N.~Itzhaki and E.~D.~Kovetz,
``Inflection Point Inflation and Time Dependent Potentials in String Theory,''
  JHEP {\bf 0710}, 054 (2007)
  [arXiv:0708.2798 [hep-th]].  %%CITATION = ARXIV:0708.2798;%%




%\cite{Badziak:2008yg}
\bibitem{Badziak:2008yg}
  M.~Badziak and M.~Olechowski,
``Volume modulus inflation and a low scale of SUSY breaking,''
  arXiv:0802.1014 [hep-th].
  %%CITATION = ARXIV:0802.1014;%%

\bibitem{08040863}
  B.~S.~Acharya, P.~Kumar, K.~Bobkov, G.~Kane, J.~Shao and S.~Watson,
  %``Non-thermal Dark Matter and the Moduli Problem in String Frameworks,''
  arXiv:0804.0863 [hep-ph].
  %%CITATION = ARXIV:0804.0863;%%

      %\cite{Holman:1984yj}
\bibitem{Holman:1984yj}
  R.~Holman, P.~Ramond \& G.~G.~Ross,
``Supersymmetric Inflationary Cosmology,''
  Phys.\ Lett.\  B {\bf 137}, 343 (1984).
  %%CITATION = PHLTA,B137,343;%%


%\cite{Linde:2007jn}
\bibitem{Linde:2007jn}
  A.~Linde and A.~Westphal,
 ``Accidental Inflation in String Theory,''
  JCAP {\bf 0803}, 005 (2008)
  [arXiv:0712.1610 [hep-th]].
  %%CITATION = JCAPA,0803,005;%%

  %\cite{Underwood:2008dh}
\bibitem{Underwood:2008dh}
  B.~Underwood,
``Brane Inflation is Attractive,''
  arXiv:0802.2117 [hep-th].
  %%CITATION = ARXIV:0802.2117;%%


\bibitem{bs}
R.~Brustein and P.~J.~Steinhardt,
``Challenges for superstring cosmology,''
  Phys.\ Lett.\  B {\bf 302} (1993) 196
  [arXiv:hep-th/9212049].
  %%CITATION = PHLTA,B302,196;%%




\bibitem{grqc9711068}
  E.~J.~Copeland, A.~R.~Liddle and D.~Wands,
``Exponential potentials and cosmological scaling solutions,''
  Phys.\ Rev.\  D {\bf 57}, 4686 (1998)
  [arXiv:gr-qc/9711068].
  %%CITATION = PHRVA,D57,4686;%%


\bibitem{astroph9711102}
  P.~G.~Ferreira and M.~Joyce,
``Cosmology with a Primordial Scaling Field,''
  Phys.\ Rev.\  D {\bf 58}, 023503 (1998)
  [arXiv:astro-ph/9711102].
  %%CITATION = PHRVA,D58,023503;%%

\bibitem{hepph0506045}
  T.~Barreiro, B.~de Carlos, E.~Copeland and N.~J.~Nunes,
``Moduli evolution in the presence of flux compactifications,''
  Phys.\ Rev.\  D {\bf 72}, 106004 (2005)
  [arXiv:hep-ph/0506045].
  %%CITATION = PHRVA,D72,106004;%%

\bibitem{Wetterich88}
  C.~Wetterich,
``Cosmology and the Fate of Dilatation Symmetry,''
  Nucl.\ Phys.\  B {\bf 302}, 668 (1988).
  %%CITATION = NUPHA,B302,668;%%

\bibitem{overshoot}
N.~Kaloper and K.~A.~Olive,
``Dilatons in string cosmology,''
  Astropart.\ Phys.\  {\bf 1} (1993) 185.
  %%CITATION = APHYE,1,185;%%
A.~Albrecht, C.~P.~Burgess, F.~Ravndal and C.~Skordis,
 ``Natural quintessence and large extra dimensions,''
  Phys.\ Rev.\  D {\bf 65} (2002) 123507
  [arXiv:astro-ph/0107573].
  %%CITATION = PHRVA,D65,123507;%%

\bibitem{hepth0408160}
  R.~Brustein, S.~P.~de Alwis and P.~Martens,
``Cosmological stabilization of moduli with steep potentials,''
  Phys.\ Rev.\  D {\bf 70}, 126012 (2004)
  [arXiv:hep-th/0408160].
  %%CITATION = PHRVA,D70,126012;%%
  I.~Ben-Dayan, R.~Brustein and S.~P.~de Alwis,
  %``Models of Modular Inflation and Their Phenomenological Consequences,''
  arXiv:0802.3160 [hep-th].
  %%CITATION = ARXIV:0802.3160;%%

\bibitem{init}
N.~Kaloper, J.~Rahmfeld and L.~Sorbo,
``Moduli entrapment with primordial black holes,''
  Phys.\ Lett.\  B {\bf 606} (2005) 234
  [arXiv:hep-th/0409226].
  %%CITATION = PHLTA,B606,234;%%
A.~Berndsen, T.~Biswas and J.~M.~Cline,
``Moduli stabilization in brane gas cosmology with superpotentials,''
  JCAP {\bf 0508} (2005) 012
  [arXiv:hep-th/0505151].
  %%CITATION = JCAPA,0508,012;%%
N.~Itzhaki and E.~D.~Kovetz,
``Inflection Point Inflation and Time Dependent Potentials in String
  Theory,''
  arXiv:0708.2798 [hep-th].
  %%CITATION = ARXIV:0708.2798;%%

\bibitem{hepth0505098}
  D.~A.~Easson and M.~Trodden,
  %``Moduli stabilization and inflation using wrapped branes,''
  Phys.\ Rev.\  D {\bf 72} (2005) 026002
  [arXiv:hep-th/0505098].
  %%CITATION = PHRVA,D72,026002;%%


\bibitem{curvaton1}  A.~D.~Linde and V.~Mukhanov,  ``Nongaussian isocurvature
perturbations from inflation,''  Phys.\ Rev.\ D \textbf{56}, 535 (1997)
[arXiv:astro-ph/9610219];  K.~Enqvist and M.~S.~Sloth,
``Adiabatic CMB perturbations in pre big bang string cosmology,''
  Nucl.\ Phys.\  B {\bf 626}, 395 (2002)
  [arXiv:hep-ph/0109214]; D.~H.~Lyth and D.~Wands, ``Generating the curvature
perturbation without an inflaton,''  Phys.\ Lett.\ B \textbf{524}, 5 (2002)
[arXiv:hep-ph/0110002]; T.~Moroi and T.~Takahashi, ``Effects of cosmological
moduli fields on cosmic microwave background,''  Phys.\ Lett.\ B \textbf{522}%
, 215 (2001)  [Erratum-ibid.\ B \textbf{539}, 303 (2002)]
[arXiv:hep-ph/0110096].
  %%CITATION = JCAPA,0604,009;%%


\bibitem{curvaton2} G.~Dvali, A.~Gruzinov and M.~Zaldarriaga, ``A new mechanism
for generating density perturbations from inflation,'' Phys.\ Rev.\ D
\textbf{69}, 023505 (2004) [arXiv:astro-ph/0303591]; L.~Kofman, ``Probing
string theory with modulated cosmological fluctuations,''
arXiv:astro-ph/0303614; A.~Mazumdar and M.~Postma,
  ``Evolution of primordial perturbations and a fluctuating decay
rate,''
   Phys.\ Lett.\ B {\bf 573}, 5 (2003)
   [Erratum-ibid.\ B {\bf 585}, 295 (2004)]
   [arXiv:astro-ph/0306509]; F.~Bernardeau, L.~Kofman and J.~P.~Uzan, ``Modulated
fluctuations from hybrid inflation,''  Phys.\ Rev.\ D \textbf{70}, 083004
(2004)  [arXiv:astro-ph/0403315];  D.~H.~Lyth, ``Generating the curvature
perturbation at the end of inflation,''  arXiv:astro-ph/0510443;
T.~Matsuda,
 ``Modulated Inflation,''
 arXiv:0801.2648 [hep-ph];
 T.~Suyama and M.~Yamaguchi,
``Non-Gaussianity in the modulated reheating scenario,''
 Phys.\ Rev.\  D {\bf 77}, 023505 (2008)
 [arXiv:0709.2545 [astro-ph]]; M.~Sasaki, ``Multi-brid inflaton and non-Gaussianity,'' in preparation.


  \bibitem{bhk}
M.~Berg, M.~Haack and B.~Kors,
  ``String loop corrections to Kaehler potentials in orientifolds,''
  JHEP {\bf 0511} (2005) 030
  [arXiv:hep-th/0508043];
  %%CITATION = JHEPA,0511,030;%%
M.~Berg, M.~Haack and E.~Pajer,
  ``Jumping Through Loops: On Soft Terms from Large Volume Compactifications,''
  JHEP {\bf 0709} (2007) 031
  [arXiv:0704.0737 [hep-th]];
  %%CITATION = JHEPA,0709,031;%%
M.~Cicoli, J.~P.~Conlon and F.~Quevedo,
``Systematics of String Loop Corrections in Type IIB Calabi-Yau Flux
Compactifications,''
  JHEP {\bf 0801}, 052 (2008)
  [arXiv:0708.1873 [hep-th]].
  %%CITATION = ARXIV:0708.1873;%%
  M.~Cicoli, J.~P.~Conlon and F.~Quevedo,
  %``General Analysis of LARGE Volume Scenarios with String Loop Moduli
  %Stabilisation,''
  arXiv:0805.1029 [hep-th].
  %%CITATION = ARXIV:0805.1029;%%


%\cite{Yadav:2007yy}
\bibitem{Yadav:2007yy}
  A.~P.~S.~Yadav and B.~D.~Wandelt,
``Detection of primordial non-Gaussianity (fNL) in the WMAP 3-year data at
above 99.5\% confidence,''
  arXiv:0712.1148 [astro-ph].
  %%CITATION = ARXIV:0712.1148;%%

  %\cite{Hikage:2008gy}
\bibitem{Hikage:2008gy}
  C.~Hikage, T.~Matsubara, P.~Coles, M.~Liguori, F.~K.~Hansen and S.~Matarrese,
``Primordial Non-Gaussianity from Minkowski Functionals of the WMAP
Temperature Anisotropies,''
  arXiv:0802.3677 [astro-ph].
  %%CITATION = ARXIV:0802.3677;%%

  %\cite{McEwen:2008kv}
\bibitem{McEwen:2008kv}
  J.~D.~McEwen, M.~P.~Hobson, A.~N.~Lasenby and D.~J.~Mortlock,
``A high-significance detection of non-Gaussianity in the WMAP 5-year data
using directional spherical wavelets,''
  arXiv:0803.2157 [astro-ph].
  %%CITATION = ARXIV:0803.2157;%%

  %\cite{Komatsu:2008hk}
\bibitem{Komatsu:2008hk}
  E.~Komatsu {\it et al.}  [WMAP Collaboration],
``Five-Year Wilkinson Microwave Anisotropy Probe (WMAP)
Observations: Cosmological Interpretation,''
  arXiv:0803.0547 [astro-ph].
  %%CITATION = ARXIV:0803.0547;%%

   \bibitem{cq}
 J.~P.~Conlon and F.~Quevedo,
 ``Astrophysical and Cosmological Implications of Large Volume String
Compactifications,''
  JCAP {\bf 0708}, 019 (2007)
  [arXiv:0705.3460 [hep-ph]].
  %%CITATION = ARXIV:0705.3460;%%

\bibitem{thermal}
D.~H.~Lyth and E.~D.~Stewart,
``Thermal Inflation And The Moduli Problem,''
  Phys.\ Rev.\  D {\bf 53} (1996) 1784
  [arXiv:hep-ph/9510204].
  %%CITATION = PHRVA,D53,1784;%%



\end{thebibliography}
\end{document}